\documentclass[aps,prb,twocolumn,groupedaddress,showpacs]{revtex4}

\usepackage{amsfonts}
\usepackage{amsmath}
\usepackage{graphicx}

\newcommand{\tr}{\mbox{tr}\:}
\newcommand{\eq}[1]{Eq.~(\ref{#1})}
\newcommand{\lso}{l_{\rm so}}
\newcommand{\lH}{l_{H}}

\newcommand{\brak}[1]{\mbox{$ \langle #1  \rangle$}}

\begin{document}

\title{Crossover from weak localization to weak antilocalization
in a disordered microbridge}

\author{M.\ G.\ A.\ Crawford}
\author{P.\ W.\ Brouwer}
\affiliation{Laboratory of Atomic and Solid State Physics, Cornell
University, Ithaca, NY 14853-2501}
\author{C.\ W.\ J.\ Beenakker}
\affiliation{Instituut-Lorentz, Leiden University, P.O. Box 9506, 2300 RA
Leiden, The Netherlands}

\date{\today}

\begin{abstract}
We calculate the weak localization correction in the double crossover to
broken time-reversal and spin-rotational symmetry for a disordered 
microbridge or a short disordered wire using a scattering-matrix approach.
Whereas the correction has universal limiting values in the three basic
symmetry classes, the functional form of the magnetoconductance is
affected by eventual non-homogeneities in the microbridge.

\end{abstract}



\maketitle

Interference of time-reversed paths causes a small negative quantum
correction to the conductance of a disordered metal termed the weak
localization.\cite{Anderson79,Gorkov79,bergmann84,LeeRamakrishnan}
This correction is suppressed by a time-reversal symmetry breaking
magnetic field, whereas in the presence of strong spin-orbit scattering,
the sign of the correction is reversed.\cite{hikami80} In that case,
the interference correction is known as weak antilocalization.

In a wire geometry at zero temperature, 
the weak localization correction takes a 
particularly simple and universal form,\cite{BeenakkerReview}
\begin{equation}
  \delta G = \frac{2 e^2 (\beta - 2)}{3 \beta h},
  \label{ago}
\end{equation}
where the symmetry parameter $\beta$ denotes the appropriate symmetry
class: In the presence of an applied magnetic field, $\beta=2$, and without
a magnetic field, $\beta=4$ or 1 with or without strong spin-orbit
scattering, respectively. Equation (\ref{ago}) was obtained using
random-matrix theory,\cite{mello88,mello91,macedo92} and diagrammatic
perturbation theory,\cite{LeeRamakrishnan,mello91} and is valid if the
length $L$ of the wire is much smaller than the localization length $\xi$
and the dephasing length $L_{\phi}$, but much larger than the mean free
path $l$. The validity of Eq.\ (\ref{ago}) extends to the
case when sample
parameters are non-homogeneous, e.g., for wires of varying cross 
section, mean free path, or electron density.

For wires with weak spin-orbit scattering, a crossover between
weak localization and weak antilocalization takes place when
the spin-orbit scattering length $l_{\rm so}$ becomes comparable
to $L$ or $L_{\phi}$ (whichever is smaller). Experimentally, this
crossover regime has been well studied in wires with length
$L \gg L_{\phi}$.\cite{kurdak92,Moon,Gougam} In this regime,
weak (anti)localization takes the form of a small correction to the
conductivity of the wire, rather than of a correction to the conductance.
Theoretically, the weak localization to weak antilocalization
crossover in the regime $L \gg L_{\phi}$ has been considered in Refs.\
\onlinecite{aa,santhanam84b,santhanam87b} using diagrammatic perturbation
theory. The opposite regime $L \ll L_{\phi}$, where the universal
correction (\ref{ago}) to the conductance $G$ can be observed, would
be relevant for relatively short high-purity metal wires,\cite{Pothier}
or disordered microbridges.

The goal of this paper is threefold: (i) to generalize the random-matrix
methods for quantum wires to the crossover between weak localization
and weak antilocalization, thus extending the equivalence of the
two methods to the interpolation between the three symmetry
classes, (ii) to find an explicit expression for $\delta G$ for $L \ll 
L_{\phi}$, and (iii) to extend the theory for the crossover
regime to the case of non-homogeneous wires, for which
the electron density, impurity concentration, or cross section
varies along the sample. In this case, both the crossover scale
and the functional form of $\delta G$ in the crossover are affected by 
non-homogeneities. The fact that the crossover scale,
characterized by the spin-orbit length $\lso$ and the
magnetic length $\lH$, is non-universal is well known, both
for homogeneous and for non-homogeneous
microbridges.\cite{BeenakkerVanHouten} 
Our finding that the functional form of the crossover is affected
by the non-homogeneity is markedly different from crossovers 
between the three basic symmetry classes in 
quantum dots, where the functional forms are universal and given
by random-matrix theory.\cite{BeenakkerReview} For homogeneous wires,
$\delta G$ is a universal function of $L/\lso$ and $L/\lH$.

The main assumption underlying our calculations is that the wire
width $W \ll L$, {\em i.e.\ } quasi one-dimensionality.
We also assume that the wire is well in the diffusive regime, 
$l \ll L, \lso, \lH \ll \xi$, where $l$ is the elastic mean free
path, and, for a non-homogeneous microbridge, that the 
number of propagating channels at
the Fermi level $N$ has only one minimum along the wire
(excluding the possibility of a ``cavity'').
We first discuss our calculations for homogeneous wires;
the case of non-homogeneous samples is discussed at the end of this 
paper.

Starting point of our calculation is a random-matrix model similar
to that used by Dorokhov.\cite{Dorokhov88} A disordered wire 
with $N$ propagating channels at the Fermi level is
modeled by $N$ one-dimensional channels and periodically
inserted scatterers that scatter within and between the channels.
The electronic wavefunction is represented by a $2N$-component
vector of spinors. 
The $2N$ components of the wavefunction refer to the transverse 
channel and to the left/right mover index. 
Linearizing the kinetic energy in each of the channels, the
Hamiltonian $H$ takes the form of a differential
operator with respect to the coordinate $x$ along the wire
and a $2N$-dimensional
quaternion matrix with respect to the channel and left/right mover
indices and spinor degree of freedom,
\begin{equation}
  H = -i \sigma_0 \otimes \tau_3 \otimes \openone_{N}
  \frac{\partial}{\partial x}
  + \sum_{j} V_j \delta(x-j a),
  \label{eq:Ham}
\end{equation}
with $\sigma_0$ the $2\times 2$ unit matrix for the spinor
degree of freedom, $\tau_3$ the Pauli matrix in 
left-mover/right-mover grading, $\openone_{N}$ the $N \times N$ unit
matrix in the channel grading, 
$V_j$ a Hermitian $2N \times 2N$ quaternion matrix 
representing the $j$th scatterer along the wire, and $a$ the distance 
between scatterers. 
A quaternion is a $2 \times 2$ matrix acting in the spinor
grading with special rules for
transposition and complex conjugation \cite{Mehta}:
The ``dual'' $X^{\rm R}$ of a quaternion matrix is
$X^{\rm R} = \sigma_2 X^{\rm T} \sigma_2$; the quaternion complex
conjugate is defined as $X^* = (X^{\dagger})^{\rm R}$.
We have chosen units such that the Fermi velocity
is one. A model
similar to Eq.\ (\ref{eq:Ham}) has been used in Ref.\
\onlinecite{brouwer02a} to study weak localization in 
unconventional superconducting wires.
 
The ensemble-averaged conductance $\langle G \rangle$ 
of the wire is given by the Landauer formula,
\begin{equation}
  \langle G \rangle = 
  \frac{e^2}{h} g, \quad g = \brak{\tr (1 - r^{\dagger} r)},
\end{equation}
where $r$ is the $N \times N$ quaternion reflection matrix of the wire.
To calculate $r$, we start from a wire of zero length
and add slices of length $a$ at the wire's ends. The
scattering matrix of the $j$th scatterer is
\begin{equation}
S_{j} = \left ( \begin{array}{cc}
t_{j} & r'_{j} \\
r_{j} & t'_{j}
\end{array} \right ) = 
\frac{2i-V_{j}}{2i+V_{j}}.
  \label{eq:SV}
\end{equation}
Hence, if a scatterer is added at the lead end of the wire, the
new reflection matrix of the wire is calculated according to the
composition rule
\begin{equation}
  r \rightarrow r_{j} + t'_{j} r (1-r'_{j}r)^{-1} t_{j}.
\label{cud}
\end{equation}
(A similar composition rule, involving both transmission
and reflection matrices of the disordered wire, 
applies if a scatterer is added at the
far end of the wire.\cite{BeenakkerReview})

In left-mover/right-mover grading, the potential $V_j$ is
parameterized as
\begin{equation}
  V = \left( \begin{array}{cc} v_{LL} & v_{LR} \\
  v_{RL} & v_{RR} \end{array} \right),
\end{equation}
where $v_{LL}$, $v_{LR}$, $v_{RL}$, and $v_{RR}$ are
$N \times N$ quaternion matrices,
\begin{subequations} \label{eq:v}
\begin{eqnarray}
\lefteqn{ v_{LL}(\alpha_f,\eta_f) = v_{RR}^*(\alpha_f,-\eta_f) } \\
  &=& \sqrt{\frac{a}{l_f N}} \left[
  (u_{f}^{0} + \eta_{f} x_{f} ) \otimes \sigma_{0} +
i \alpha_{f} \sum_{\mu=1}^{3} u_{f}^{\mu} \otimes \sigma_{\mu} \right],
\nonumber \\
\lefteqn{ v_{LR}(\alpha_b,\eta_b) = v_{RL}^{\dag}(\alpha_b,\eta_b) } \\
 &=& \sqrt{\frac{a}{l (N+1)}} \left[
  (u_{b}^{0} + \eta_{b} x_{b} ) \otimes \sigma_{0} +
i \alpha_{b}  \sum_{\mu=1}^{3} u_{b}^{\mu} \otimes \sigma_{\mu} \right].
\nonumber
\end{eqnarray}
\end{subequations}
In Eq.\ (\ref{eq:v}),
$u_{f}^{0}$ and $x_{f}$ are random Hermitian $N \times N$ matrices,
$u_{f}^{\mu}$, $\mu=1,2,3$, is a random anti-Hermitian matrix,
$u_{b}^{0}$ is a random symmetric matrix, and $u_{b}^{\mu}$, 
$\mu=1,2,3$ and $x_b$ are random antisymmetric matrices. 
All of these random matrices have independent and Gaussian distributions
with zero mean and unit variance. (Variances are specified for the
off-diagonal elements; diagonal elements have double variance for
symmetric matrices and are zero for antisymmetric matrices.) 
The parameters $\alpha_b$ and $\alpha_f$ describe the strength
of the breaking of spin-rotational symmetry; The
parameters $\eta_b$ and $\eta_f$ describe the strength of the breaking
of time-reversal symmetry.  Finally, $l_f$ is the elastic mean free path
for forward scattering and $l$ is the transport mean free path.

To find the conductance of the wire we calculate the change of
$g$ if one scatterer is added to the wire. To this end, we expand the
scattering matrix $S_j$ of Eq.\ (\ref{eq:SV}) in powers of $V_j$, use
the composition rule (\ref{cud}), and calculate the Gaussian average over
the potential $V_j$.  In the limit $a \ll l$ of weak disorder we thus find
\begin{equation}
-2 N l \frac{\partial}{\partial L} g = g^2 - h_{0} + 3 h_{1}.
\label{eq:rdiff}
\end{equation}
We abbreviated
\begin{subequations} \label{eq:hdef}
\begin{eqnarray}
  h_0 &=& \langle \tr (1 - r^{\dagger} r)
  (1 - r^{*} r^{\rm R}) \rangle,
  \\
  h_1 &=& \frac{1}{3} \sum_{\mu=1}^{3}
  \langle \tr (1 - r^{\dagger} r) \sigma_{\mu}
  (1 - r^{*} r^{\rm R}) \sigma_{\mu} \rangle,
\end{eqnarray}
\end{subequations}
and omitted terms that vanish in the diffusive regime
$l \ll L, \lso, \lH \ll N l$.
The subscripts $0$ and $1$ refer
to singlet and triplet contributions, respectively.

To leading order in $N$, Eq.\ (\ref{eq:rdiff}) can be solved
without the interference corrections $h_0$ 
and $h_1$, with the result
\begin{equation} 
  g =  \frac{2 N l}{L} + O(1),
\label{giraffe}
\end{equation}
corresponding to the Drude law for the conductance. The $O(1)$ 
correction in Eq.\ (\ref{giraffe}) gives the weak localization 
correction $\delta g$, which we now compute.

To find the weak localization correction, we need to
calculate $h_0$ and $h_1$. Proceeding as before, we find that
the $L$-dependence of $h_m$, $m=0,1$ is governed by the evolution
equation
\begin{equation}
  2 N l \frac{\partial h_m}{\partial L}
  = -2 \left ( \frac{2N l}{L} + k_{m} \right ) h_{m}
    + \frac{8 N^{2} l^{2}}{L^{2}} ,\ \ m=0,1,
  \label{eq:hevol}
\end{equation}
where we abbreviated
\begin{eqnarray}
  k_0 = \langle \tr (1-r^* r) \rangle,\ \
  k_1 = \frac{1}{3} \sum_{\mu=1}^{3} \langle \tr (1-r^* \sigma_{\mu} r
  \sigma_{\mu}) \rangle.
\end{eqnarray}
Evolution equations for $k_0$ and $k_1$ are obtained similarly
and read
\begin{subequations} \label{eq:kevol}
\begin{eqnarray}
  2 N l \frac{\partial k_0}{\partial L}
  &=& \left( \frac{2 N l}{\lH} \right)^2 - k_0^2,\\
  2 N l \frac{\partial k_1}{\partial L}
  &=& \left( \frac{2 N l}{\lH'} \right)^2 - k_1^2,
\end{eqnarray}
\end{subequations}
where the length scales $\lH$ and $\lH'$ are defined in terms
of the parameters of the random-matrix model (\ref{eq:v}),
\begin{subequations}
\begin{eqnarray}
  \lH^{-2} &=& 2(l^{-2} \eta_{b}^{2}
              + l^{-1} l_{f}^{-1} \eta_{f}^{2} ),\\
  \lso^{-2} &=& 6 (l^{-2} \alpha_{b}^{2}
              + l^{-1} l_{f}^{-1} \alpha_{f}^{2} ), \\
  (\lH')^{-2} &=& \lH^{-2} + {\textstyle \frac{4}{3}} \lso^{-2}.
\end{eqnarray}
\end{subequations}

Equations (\ref{eq:hevol}) and (\ref{eq:kevol}) have the solution
\begin{subequations} \label{eq:h}
\begin{eqnarray}
  k_0 &=& \frac{2Nl}{\lH} \coth  \frac{L}{\lH},\\
  h_0 &=& \frac{2Nl}{L} \left(1 + \frac{\lH}{L} \coth \frac{L}{\lH} -
  \coth^2 \frac{L}{\lH} \right). 
\end{eqnarray}
\end{subequations}
Expressions for $k_1$ and $h_1$ are obtained from
Eq.\ (\ref{eq:h}) after the substitution $\lH \to \lH'$.
Substitution of $h_0$ and $h_1$ into
Eq.\ (\ref{eq:rdiff}) then allows for the calculation of the
weak-localization correction to the conductance,
\begin{equation}
  \delta g =  
  \frac{\lH}{L} \coth \frac{L}{\lH} - \frac{\lH^2}{L^2}
 - 3 \left ( \frac{\lH'}{L} \coth \frac{L}{\lH'} - \frac{(\lH')^{2}}{L^2}
   \right ).
  \label{job}
\end{equation}
At zero magnetic field, Eq.\ (\ref{job}) simplifies to
\begin{equation}
  \delta g =
  \frac{1}{3} + \frac{9 \lso^{2}}{4 L^{2}} - 
  \frac{3 \lso\sqrt{3} }{2 L} \coth \frac{2L}{\lso \sqrt{3}}.
\label{toe}
\end{equation}

Equation (\ref{toe}) reproduces the limits $\delta G = 
- 2 e^2/3h$ without spin-orbit scattering and $\delta G 
= e^2/3h$ with strong spin-orbit scattering. 
Without spin-orbit scattering, Eq.\ (\ref{job}) agrees with
the weak localization correction calculated
in Ref.\ \onlinecite{altshuler84}.
For large magnetic fields, $L \gg \lH$, Eq.\ (\ref{job}) simplifies
to
\begin{equation}
  \delta g = \frac{1}{L}
  \left(\lH - 3 (\lH^{-2} +
  {\textstyle \frac{4}{3}}
  \lso^{-2})^{-1/2} \right),
  \label{eq:llarge}
\end{equation}
which has the same functional form as the weak
localization obtained using diagrammatic perturbation
theory.\cite{altshuler81,aa,santhanam84b,santhanam87b}
Comparison of Eq.\ (\ref{eq:llarge}) and Refs.\
\onlinecite{altshuler81,aa,santhanam84b,santhanam87b} allows us to identify
$\lso$ as the spin-orbit length, and, for a channel (with width
$W \gg l$) in a two-dimensional electron gas in a
perpendicular magnetic field $B$,
\begin{equation}
  \lH^2 = 3 (\hbar/W B e)^2.
\label{all}
\end{equation}
The case of a cylindrical wire of radius $R \gg l$ 
and magnetic field perpendicular to the wire is obtained by
the substitution $W^2 \to 3 R^2/2$. 
For $l > W$ (or $l > R$) the crossover length $\lH$ has a more 
complicated $l$-dependent expression.\cite{BvH}

\begin{figure}
\includegraphics[width=2.5in] {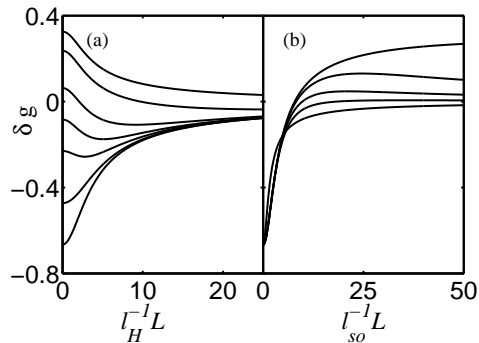} \vspace{-0.5cm}
\caption{The weak localization correction $\delta g$ plotted (a)
as a function of the magnetic field strength (characterized by the
dimensionless ratio $\lH^{-1}L$) for fixed value of the spin-orbit
scattering rate (characterized by $\lso^{-1} L$). From bottom to
top, the curves
correspond to $L/\lso = 0.1$, $2$, $4$, $6$, $10$, $30$, and $\infty$.
(b) as a function of length $L$ for fixed $\lH^{-1}\lso$.
From bottom to top, the curves correspond to 
$\lH^{-1}\lso = 2$, $0.3$, $0.2$, $0.1$, and $0$.
\label{weak}}
\end{figure}

Figure \ref{weak}(a) shows $\delta g$ as a function of the magnetic 
field for several values of the spin-orbit coupling. In Fig.\
\ref{weak}(b) we show $\delta g$ as a function of $\lso^{-1}L$ for several
values of the magnetic field.

\begin{figure}
\includegraphics[width=2.5in] {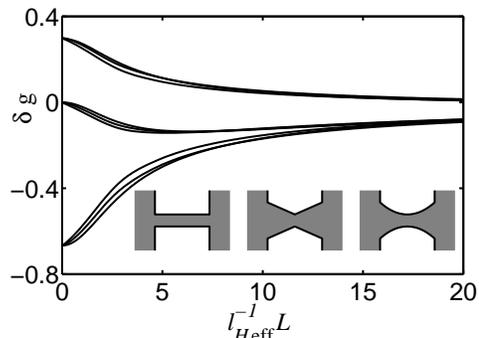} \vspace{-0.5cm}
\caption{The weak localization correction $\delta g$ as a function of
the magnetic field strength 
for three different shapes of a disordered microbridge (channels
in a two-dimensional electron gas). 
The three different shapes are characterized by
$s(x)=1$, $s(x)=1+ 4 |2 x/L|$
and $s(x)=1+4 (2x/L)^{2}$, $-L/2 < x < L/2$, cf.\ Eq.\
(\protect\ref{eq:sdef}),
as shown in the inset. The three groups of curves correspond to
strong, intermediate and weak spin-orbit scattering from top to bottom,
with $\lso$ in the intermediate case chosen for each case to render
the same correction as $\lH^{-1} \rightarrow 0$. The magnetic field
strength is measured in terms of the effective
magnetic length $l_{H,{\rm eff}}$, cf.\ \eq{eq:23}.
\label{sweep}}
\end{figure}

We now turn to a description of the weak localization correction in
a non-homogeneous microbridge. Examples of non-homogeneous 
microbridges with varying widths are shown in the inset
of Fig.\ \ref{sweep}. If the wire cross section or the electron density
vary with the coordinate $x$ along the wire, the number of propagating 
channels at the Fermi level $N$ also varies with $x$. We assume
that $N(x)$ has a minimum for $x=0$ and that $dN/dx > 0$ 
($dN/dx < 0$) for all $x > 0$ ($x < 0$). Further, $x$-dependence
of the impurity concentration, the smoothness of the boundary,
the shape of the cross section, etc., causes an $x$-dependence 
of the length scales $l$, $\lH$, and $\lso$.

The
reflection matrix of the wire is constructed by building the wire 
from thin
slices, starting at the narrowest point $x=0$. This
way, the number of channels in the slices added
to both ends of the wire can increase, but not decrease.
For the construction of an evolution equation for the
conductance $g$ and for the auxiliary functions $h_0$, $h_1$,
$k_0$, and $k_1$, we distinguish between two types of
added slices: A thin slice that contains a scattering site but
for which the number of channels remains constant, and a thin
slice without scatterer in which $N$ increases by unity.
Addition of a slice of the former type causes 
a small change in the reflection
matrix $r$, which is the same as for a quantum wire of constant
thickness, see Eq.\ (\ref{cud}) above.
Addition of a slice for which $N$ increases by unity does not 
cause a change of the conductance
$g$ or of the auxiliary functions $h_0$, $h_1$, $k_0$, or $k_1$, 
as can seen by inspecting
the cases $x > 0$ and $x < 0$ separately: For $x > 0$, 
an increase of $N$ does not cause additional 
reflection, and hence does not affect the reflection matrix
$r$; for $x < 0$, an increment
of $N$ changes the dimension of the reflection matrix $r$ by one,
\begin{equation}
  r \to \left( \begin{array}{cc} r & 0 \\ 0 & 1 \end{array}
  \right), \label{cud2}
\end{equation}
but does not change the conductance $g$, or the
functions $h_0$, $h_1$, $k_0$, or $k_1$.
Combining the two types of slices,
we conclude that the only effect of the 
$x$-dependence of $N$ and $l$ is indirect, through the explicit
appearance of $N$ and $l$ in statistics of the scattering
matrix of the added slice, see Eq.\ (\ref{eq:v}).
In the diffusive regime,
$N(x)$ and $l(x)$ only appear in the combination
\begin{equation}
  s(x) = {N(x) l(x)}/{N_0 l_0}, \label{eq:sdef}
\end{equation}
where $N_0$ and $l_0$ are number of propagating channels and mean 
free path at $x=0$. For large $N$ the function $s(x)$
may be considered continuous, and the evolution equations 
become differential equations which now include explicit reference
to the function $s(x)$. If the wire length $L$ is replaced by the
effective length $\bar L$,
\begin{equation}
\bar{L} = \int
\frac{d x }{s(x)},
\label{hut}
\end{equation}
the evolution equations for $g$,
$h_0$, $h_1$, $k_0$, and $k_1$ keep the same form as for
homogeneous wires, provided
we make the substitutions $N \to N_0$, $L \to \bar L$, $l
\to l_0$, 
  $\lH \to \overline{\lH} = \lH /s(x)$, and $\lso \to \overline{\lso}
= \lso/s(x)$.

The 
functional form of the leading-in-$N$ contribution to the 
conductance remains unchanged, $G = (e^2/h) (2N_0 l_0/\bar L)$. Also,
for the limiting cases of no spin-orbit scattering and strong 
spin-orbit scattering, the weak localization correction $\delta G$ 
is still given by the universal result Eq.\
(\ref{ago}).\cite{BeenakkerMelsen94} However, because of the 
$x$-dependence
of the length scales $\overline{\lH}$ and $\overline{\lso}$, 
$\delta g$ acquires an
explicit dependence on the shape of the disordered microbridge
or the non-homogeneity of the mean free path or the electron density
in the crossover region between the symmetry classes.
For a large magnetic field ($\lH^{-1}L \gg 1$), the weak-localization
correction can be found in closed form,
\begin{eqnarray}
  \delta g &=& \frac{1}{\bar{L}} \label{eq:23}
  \left(l_{H,{\rm eff}} - 3 l'_{H,{\rm eff}} \right), \\
  l_{H,{\rm eff}} &=&
  \frac{1}{\bar{L}} \int \frac{\lH (x) dx}{s(x)^2 }, \ \ 
  l'_{H,{\rm eff}} =
  \frac{1}{\bar{L}} \int \frac{\lH' (x) dx}{s(x)^2}.
  \nonumber
\end{eqnarray}
Equation (\ref{eq:23}) simplifies to \eq{eq:llarge} in the 
case of $s(x)$ constant. The same result follows if 
\eq{eq:llarge}
is interpreted as a quantum interference correction to the 
one-dimensional resistivity and $\lH$ is taken $x$-dependent.
For weaker magnetic fields with $\lH^{-1} L$ of order unity, a
numerical solution of the evolution equations is required.

In Fig.\ \ref{sweep}, we show results of a numerical solution 
of $\delta g$ for the examples $s(x)$ constant, $s(x) = 1 + 4
|2 x/L|$ and $s(x) = 1 +  4 (2 x/L)^2$, $-L/2 < x < L/2$. These
functional forms correspond to diffusive microbridges in a
two-dimensional electron gas of the form 
shown in the inset of Fig.\ \ref{sweep} with uniform impurity
concentration and mean free path $l \ll W$. 
The three sets of curves in the figure
represent strong, intermediate
and weak spin-orbit scattering, respectively.  For the intermediate
case (middle set of curves in Fig.\ \ref{sweep}), 
three different values of $\lso$ were chosen so that the
weak-localization correction $\delta g = 0$ is equal in the three 
cases for zero magnetic field. The magnetic field is
characterized by the ratio $l_{H,{\rm eff}}^{-1}L$, cf.\
\eq{eq:23}, in order to remove a spurious shape
dependence for the large-field asymptotes. While there is
no dependence on the form of the function $s(x)$ in 
the limiting cases of zero and
large magnetic fields, we observe that, indeed, $\delta g$ 
depends on the precise form of the non-homogeneity for
intermediate magnetic field strengths, although, with 
proper scaling, the difference between the results for the three
cases we considered is less than 10\%.

In conclusion, we have shown that the scattering matrix approach to
quasi one-dimensional weak localization can be used to obtain a
detailed description of the crossover between the different
universality classes. We have recovered some results known from
diagrammatic perturbation theory, and have discovered one aspect of
the problem that has not been noticed previously: The dependence of the
functional form of the crossover on non-homogeneities in the
conductor.

We thank V.\ Ambegaokar, N.\ W.\ Ashcroft, and
D. Davidovic for discussions. This work was supported by the
NSF under Grant Nos.\ DMR 0086509 and DMR 9988576, by the
Packard Foundation, by the Natural Sciences and Engineering
Research Council of Canada, and by the Dutch Science Foundation 
NWO/FOM.

\end{document}